\documentclass{ws-ijmpcs}

\begin{document}

\markboth{A. Valcarce, J. Vijande, T.F. Caram\'es}
{Doubly charmed mesons}

%
\catchline{}{}{}{}{}
%

\title{DOUBLY CHARMED MESONS}

\author{A. VALCARCE}

\address{Departamento de F{\'\i}sica Fundamental, Universidad de Salamanca\\
Plza. Merced S/N, Salamanca, E-37008, Spain\\
valcarce@usal.es}

\author{J. VIJANDE}

\address{Departamento de F\'{\i}sica At\'omica, Molecular y Nuclear,\\ Universidad de Valencia (UV)
and IFIC (UV-CSIC)\\
Valencia, E-46100, Spain \\
javier.vijande@uv.es}

\author{T.F. CARAM\'ES}

\address{Departamento de F{\'\i}sica Fundamental, Universidad de Salamanca\\
Plza. Merced S/N, Salamanca, E-37008, Spain\\
carames@usal.es}

\maketitle

\begin{history}
\received{Day Month Year}
\revised{Day Month Year}
\end{history}

\begin{abstract}
Doubly charmed mesons are studied within a quark model
framework. We solve the four-quark Schor\"dinger equation by means of 
a variational approach by using different interacting potentials.
Our results point to the existence of a stable isoscalar doubly
charmed four-quark meson with quantum numbers $J^P=1^+$.

\end{abstract}

\ccode{PACS numbers: 14.40.Rt,21.30.Fe,12.39.Jh}

The potentiality of the quark model for hadron physics in the low--energy regime
became first manifest when it was used to classify the known hadron states. Describing
hadrons as $q\bar q$ or $qqq$ configurations, their quantum numbers were correctly explained.
This assignment was based on the comment by Gell-Mann\cite{Ge64}
introducing the notion of quark: {\it 'It is assuming that the lowest baryon
configuration ($qqq$) gives just the representations 1, 8 and 10, that
have been observed, while the lowest meson configuration ($q \bar{q}$)
similarly gives just 1~and~8'}. Since then, it has been assumed
that these are the only two configurations involved in the description
of physical hadrons. However, color confinement is also compatible with other multiquark
structures, like the tetraquark $qq\bar q\bar q$ first introduced by Jaffe\cite{Jaf77}.

The idea of unconventional quark structures is quite old
and despite decades of progress, no exotic meson has been
conclusively identified. A conclusive evidence of its existence
seems rather difficult to be obtained with states that could be described
by means of the lowest order configurations suggested by Gell-Mann,
like the scalar mesons or uncharged charmonium resonances.
In such cases, the
experimental data will hardly exclude a theoretical model\cite{Bug04}.
The study of exotic structures that cannot be accommodated in the 
scheme proposed by Gell-Mann
should be of primary experimental and theoretical interest, because
their existence would make manifest the contribution of four-quark states 
to hadron spectroscopy\cite{Fer09}. Among these exotic structures there are 
two types that could be experimentally accessible in present facilities:
charged charmonium states\cite{Car10} and doubly heavy mesons. 

Doubly heavy mesons is one of the first scenarios where the existence of bound multiquarks
was proposed, systems composed of two light quarks
and two heavy antiquarks ($nn\bar Q\bar Q$). 
Although they may be experimentally difficult to produce and
also to detect\cite{Mo96} it has been argued that for 
sufficiently large heavy quark mass they 
should be bound\cite{Zou86}. Their stability relies on the
heavy quark mass. The heavier the quark
the more effective the short-range Coulomb attraction to generate binding,
in such a way that it could play a decisive role to bind the system.
Moreover the $\bar Q \bar Q$ pair brings a small kinetic energy into the 
system contributing to stabilize it.

In this talk we revise the possible existence of stable doubly heavy
mesons within the nonrelativistic quark model. For this purpose we solve the four-body
Schr\"odinger equation based on a
variational method in terms of generalized gaussians\cite{Vij09}.
The four-quark wave function is written as a sum of outer products
of color, isospin, spin and configuration terms
\begin{equation}
    |\phi_{CISR}>= |{\rm Color}> |{\rm Isospin}>
               \left[|{\rm Spin}> \otimes| R > \right]^{J M} \, ,
\end{equation}
such that the four-quark state is a color singlet with well defined
parity, isospin and total angular momentum.
Coupling the color states of two quarks (antiquarks) can yield two possible
representations, the symmetric $6$-dimensional, $6$ ($\bar 6$),
and the antisymmetric $3$-dimensional, $\bar 3$ ($3$).
Coupling the color states of the quark pair with that of the antiquark pair
must yield a color singlet. Thus, there are only two possible color states for a
$QQ\bar q \bar q$ system,
\begin{equation}
  |{\rm Color}> = \{ | \bar 3_{12} 3_{34} > , | 6_{12} \bar 6_{34} >\}\,.
\label{pp}
\end{equation}
These states have well defined symmetry under permutations.
The spin states with such symmetry can be obtained in the following way,
\begin{equation}
  |{\rm Spin}> = |((s_1,s_2)S_{12},(s_3,s_4)S_{34})S>
                  = | (S_{12} S_{34}) S >\;.
\end{equation}
The same holds for the isospin, $|{\rm Isospin}>=|(i_3,i_4)I_{34} >$,
which applies only to the $n$-quarks, thus $I=I_{34}$.

To describe the spatial part of the wave function we choose for 
convenience the Jacobi coordinates,
\begin{eqnarray}
\vec{x_1} & =&  \vec r_1 - \vec r_2 \, , \cr
\vec{x_2} & =&  \vec r_3 - \vec r_4 \, , \cr
\vec{x_3} & =&  
                 \frac{m_1\vec r_1+m_2\vec r_2}{m_1+m_2}
                      -\frac{m_3\vec r_3+m_4\vec r_4}{m_3+m_4} \, , 
\end{eqnarray}
Using these vectors, it is easy
to obtain basis functions that have well defined symmetry under permutations
of the pairs $(12)$ and $(34)$. The radial wave function is expanded 
as a linear combination of generalized gaussians
\begin{equation}
|R>=\sum_{i=1}^{n} \beta^{(i)} R^i(\vec x_1,\vec x_2,\vec x_3)=\sum_{i=1}^{n} \beta^{(i)} R^i
\label{wave}
\end{equation}
where $n$ is the number of gaussians used for each color-spin-flavor component and
$R^i$ depends on six variational parameters,
$a^i$, $b^i$, $c^i$, $d^i$, $e^i$, and $f^i$, one for each scalar quantity:
$\vec x_1^{\,2}$, $\vec x_2^{\,2}$, $\vec x_3^{\,2}$, $\vec{x}_1\cdot\vec{x}_2$, 
$\vec{x}_1\cdot\vec{x}_3$ and $\vec{x}_2\cdot\vec{x}_3$.
Once the spin, color and flavor parts are integrated out the coefficients of the radial wave function are
obtained by solving the system of linear equations
\begin{equation}
\label{funci1g}
\sum_{i} \beta^{(i)}
\, [\langle R^{(j)}|\,H\,|R^{(i)}
\rangle - E\,\langle
R^{(j)}|R^{(i)}\rangle ] = 0
\qquad \qquad \forall \, j\, ,
\end{equation}
where the eigenvalues are obtained by a minimization procedure.

Assuming nonrelativistic quantum mechanics, the
hamiltonian will be given by
\begin{equation}
H=\sum_{i=1}^4\left(m_{i}+\frac{\vec p_{i}^{\,2}}{2m_{i}}\right)+\sum_{i<j=1}^4V(\vec r_{ij}) \, ,
\label{ham}
\end{equation}
where the potential $V(\vec r_{ij})$ corresponds to an arbitrary two-body interaction. 
We have analyzed the possible
existence of four-quark bound states by using two standard quark-quark
interactions, a Bhaduri-like potential (BCN)\cite{Bha81} and
a constituent quark model considering boson
exchanges (CQC)\cite{Vij05}. Both interactions
give a reasonable description of the meson and the baryon spectroscopy,
a thoughtful requirement considering that in the tetraquarks $qq$ and $q\bar q$
interactions will contribute.

We have performed an exhaustive analysis of the
$QQ\bar q \bar q$ spectra by means of the quark models described above.
We have considered all isoscalar
and isovector states with total orbital angular momentum $L\le 1$\cite{Vij09b}.
In Table~\ref{t1} we show the results for some selected quantum numbers
obtained with the CQC model. 
Our study draws a first general conclusion: once the four--body problem
is properly solved the number of bound states is rather small if not zero. For the 
$cc\bar n \bar n$ only the $J^P=1^+ (I=0)$ is bound independently of the interacting
potential, already proposed more than twenty years ago in Ref.~\refcite{Zou86}.

We have also studied all ground states of the bottom sector
using the same interacting potentials. 
It was pointed out in the early 80's that a $QQ\bar Q'\bar Q'$ four--quark state should be stable against
dissociation into $Q\bar Q'+Q\bar Q'$ if the ratio $m_Q/m_{Q'}$ is large enough\cite{Ade82}.
We observed that all bound states become deeper than
in the charm sector and a new state, $J^P=0^+ (I=0)$ appears,
strengthening the conclusion that the larger the ratio of the quark masses
the larger the binding energy.

In connection with the interacting potential used, it has been recently
analyzed the stability of $QQ\bar n\bar n$ and
$Q\bar Q n \bar n$ systems in a simple string model\cite{Vij07b} considering
only a multiquark confining interaction given by the minimum
of a flip-flop or a butterfly potential in an attempt to discern whether
confining interactions not factorizable as two--body potentials would influence
the stability of four--quark states. The ground state of
systems made of two quarks and two antiquarks of equal masses
was found to be below the dissociation threshold.
While for the cryptoexotic $Q\bar Q n\bar n$ the binding decreases
when increasing the mass ratio $m_Q/m_n$, for the flavor exotic $QQ\bar n\bar n$
the effect of mass symmetry breaking is opposite. 
\begin{table}[ph]
\tbl{Four--quark state properties for selected quantum numbers.
All states have positive parity and total orbital angular momentum $L=0$.
Energies are given in MeV and distances in fm. See text for details.}
{\begin{tabular}{@{}c|ccccc@{}} \toprule
$(S_T,I)$                       & (0,1)          &  (1,1)          & (1,0)         & (1,0)          & (0,0) \\
Flavor                          &$cc\bar n\bar n$&$cc\bar n\bar n$&$cc\bar n\bar n$&$bb\bar n\bar n$&$bb\bar n\bar n$\\
\colrule
Energy ($E_ {4q}$)              & 3877           &  3952           & 3861          & 10395          & 10948 \\
Threshold ($T=M_1+M_2$)         & $DD\mid_S$     &  $DD^*\mid_S$   & $DD^*\mid_S$   & $BB^*\mid_S$   &  $B_1B\mid_P$\\
$\Delta_E=E_{4q} -T$            & +5             &  +15            & $-76$         & $-$217         &  $-153$ \\
\colrule
$P_{MM}$                        & 1.000          &  $-$            & $-$           & $-$            &  0.254 \\
$P_{MM^*}$                      & $-$            &  1.000          & 0.505         & 0.531          &  $-$ \\
$P_{M^*M^*}$                    & 0.000          &  0.000          & 0.495         & 0.469          &  0.746 \\
\colrule
$\langle x_1^2\rangle^{1/2}$      & 60.988         &  13.804         & 0.787         & 0.684          &  0.740 \\
$\langle x_2^2\rangle^{1/2}$      & 60.988         &  13.687         & 0.590         & 0.336          &  0.542 \\
$\langle x_3^2\rangle^{1/2}$      & 0.433          &  0.617          & 0.515         & 0.503          &  0.763 \\
$RMS_{4q}$                      & 30.492         &  6.856          & 0.363         & 0.217          &  0.330 \\
\botrule
\end{tabular} \label{t1}}
\end{table}

Let us finally discuss the role played by hidden-color configurations,
color singlets built by nonsinglet constituents. Besides Eq.~(\ref{pp}),
there are other two different ways for coupling two quarks and two antiquarks
in a colorless state,
\begin{eqnarray}
[(q_1\bar q_3)(q_2\bar q_4)]&=&\{|1_{13}1_{24}\rangle,|8_{13}8_{24}\rangle\}\cr
[(q_1\bar q_4)(q_2\bar q_3)]&=&\{|1_{14}1_{23}\rangle,|8_{14}8_{23}\rangle\} \, .
\label{ll}
\end{eqnarray}
These two basis contain
both singlet--singlet (physical) and octect--octect (hidden--color) components.
It is possible to prove that there is a minimum value
for the octect--octect component probability of the wave function either in the
$[(Q_1\bar n_3)(Q_2\bar n_4)]$ or the $[(Q_1\bar n_4)(Q_2\bar n_3)]$ couplings:
$P_{88}^{13,24},P_{88}^{14,23}\in[1/3,2/3]$.
Does this imply an important hidden--color
component in all $QQ\bar n\bar n$ states? The answer is no\cite{PRC08}, because 
one can express any $QQ\bar n\bar n$ state in terms of the singlet--singlet
component of the $[(Q_1\bar n_3)(Q_2\bar n_4)]$ and $[(Q_1\bar n_4)(Q_2\bar n_3)]$ basis.

This discussion can be made more quantitative. Let us assume that $\{P,Q\}$ and $\{\hat P,\hat Q\}$
are the projectors associated to two orthonormal basis that are not orthogonal to each
other, i.e., $P\hat P \mid\phi\rangle \ne 0$ and $P\hat Q \mid\phi\rangle \ne 0$
for an arbitrary state $\mid\phi\rangle$.
This would be the case of the two orthonormal basis in Eq.~(\ref{ll}). 
For any arbitrary state
$\mid\Psi\rangle \, = \, P\mid\Psi\rangle \, + \, Q\mid\Psi\rangle \, ,$
the probability associated to $P$ or $\hat P$ will be given by\cite{PRC08},
\begin{eqnarray}
{\cal P}^{\mid\Psi\rangle}({[u]})&=&\frac{1}{2(1-\cos^2\alpha)}
\left[ \left\langle\Psi\mid P\hat Q \mid\Psi\right\rangle
+\left\langle\Psi\mid \hat Q P \mid\Psi\right\rangle\right] \cr
{\cal P}^{\mid\Psi\rangle}({[u']})&=&\frac{1}{2(1-\cos^2\alpha)}
\left[ \left\langle\Psi\mid \hat P Q \mid\Psi\right\rangle +
\left\langle\Psi\mid Q \hat P \mid\Psi\right\rangle\right] \, ,
\label{proeq}
\end{eqnarray}
where
$P=\left.\mid u \right\rangle \left\langle u \mid\right.$ and
$\hat P=\left.\mid u' \right\rangle \left\langle u' \mid\right.$ and
${\rm cos}\alpha=\left\langle u' \mid u \right\rangle$. For a molecular
state either ${\cal P}^{\mid\Psi\rangle}({[u]})$
or ${\cal P}^{\mid\Psi\rangle}({[u']})$ would be close to zero while
for a compact state both will be different from zero.

We show in Table~\ref{t1} the meson-meson probabilities for
some selected four--quark states according to Ref.~\refcite{PRC08}.
Unbound states converge to two isolated mesons, the lowest threshold of
the system, its root mean square radius (RMS) being
very large. In contrast, bound states have a radius smaller than the threshold and they
present probabilities different from zero for several physical states, the lowest two-meson
threshold being contained in the physical four--quark system. Such states would be called
compact in our notation. When the binding energy approaches the threshold, the probability of
a single physical channel converges to one, what we defined as a molecular state. 
The $cc\bar n\bar n$ $J^P(I)=1^+(0)$ is therefore a clear example of a real
tetraquark, the r.m.s. of all Jacobi coordinates being small and 
with a significant probability of different vectors of the Hilbert
space, pseudoscalar-vector ($MM^*$) and vector-vector ($M^*M^*$)
components.

\section*{Acknowledgments}

This work has been partially funded by the Spanish Ministerio de
Educaci\'on y Ciencia and EU FEDER under Contracts No. FPA2007-65748 
and FPA2010-21570, by Junta de Castilla y Le\'{o}n under Contract No. GR12,
and by the Spanish Consolider-Ingenio 2010 Program CPAN (CSD2007-00042),


\end{document}